\documentclass{article}
\usepackage[utf8]{inputenc}
\usepackage{amssymb}
\usepackage{amsfonts}
\usepackage{amsmath}
\usepackage{array}
\usepackage{esvect}
\usepackage{cite}
\usepackage[dvipsnames]{xcolor}
\usepackage{physics}
\usepackage{mathrsfs}
\usepackage{authblk}
\usepackage{geometry}
\usepackage{abstract}
\usepackage[english]{babel}
\usepackage[T1]{fontenc}
\usepackage[colorlinks,urlcolor=blue,citecolor=blue]{hyperref}
\def\hhref#1{\href{http://arxiv.org/abs/#1}{#1}} 
 \newcommand{\sfrac}[2]{{\textstyle\frac{#1}{#2}}}

\begin{document}

\begin{center}
{\LARGE \bf
Polynomial algebras from $su(3)$ and the generic model on the two sphere\\
}
\vspace{6mm}
{\Large F. Correa$^a$, M. del Olmo$^{b}$, I. Marquette$^c$ and  J. Negro$^b$ 
}
\\[6mm]
\noindent ${}^a${\em 
Instituto de Ciencias F\'isicas y Matem\'aticas\\
Universidad Austral de Chile, Casilla 567, Valdivia, Chile}\\[3mm]
\noindent ${}^b${\em
Departamento de F\'{\i}sica Te\'orica, At\'omica y
\'Optica, \\Universidad de Valladolid,  47011 Valladolid, Spain }\\[3mm]
\noindent ${}^c${\em
School of Mathematics and Physics, \\
The University of Queensland
Brisbane,  QLD 4072, Australia}
\vspace{12mm}
\end{center}

\begin{abstract}
Construction of superintegrable systems based on Lie algebras have been introduced over the years. However, these approaches depend on explicit realisations, for instance as a differential operators, of the underlying Lie algebra. This is also the case for the construction of their related symmetry algebra which take usually the form of a finitely generated quadratic algebra. These algebras often display structure constants which depend on the central elements and in particular on the Hamiltonian. In this paper, we develop a new approach reexamining the case of the generic superintegrable systems on the 2-sphere for which a symmetry algebra is known to be the Racah algebra $R(3)$. Such a model is related to the 59 $2D$ superintegrable systems on conformally flat spaces and their 12 equivalence classes. We demonstrate that using further polynomials of degree 2,3 and 4 in the enveloping algebra of $su(3)$ one can generate an algebra based only on abstract commutation relations of $su(3)$ Lie algebra without explicit constraints on the representations or realisations. This construction relies on the maximal Abelian subalgebra, also called MASA, which are the Cartan generators and their commutant. We obtain a new 6-dimensional cubic algebra where the structure constant are integer numbers which reduce from a quartic algebra for which the structure constant depend on the Cartan generator and the Casimir invariant. We also present other form of the symmetry algebra using the quadratic and cubic Casimir invariants of $su(3)$. It reduces as the known quadratic Racah algebra $R(3)$ only when using an explicit realization. This algebraic structure describe the symmetry of the generic superintegrable systems on the 2 sphere. We also present a contraction to another 6-dimensional cubic algebra which would corresponding to the symmetry algebra of a Smorodinsky-Winternitz model.
\end{abstract}

\section{Introduction} \label{sec1}

In the recent years, superintegrable systems with quadratic integrals of motion on conformally flat space and various generalizations have been classified \cite{mil13}. These models display various important properties in classical and quantum mechanics such as multiseparability of the related Hamilton-Jacobi or Schr\"odinger equations. They are also exactly solvable and their energy spectrum degeneration can be explained via the symmetry algebra. In classical mechanics, all the bounded trajectories are closed with periodic motion. They can be build via different approaches, for examples, recurrence relations of orthogonal polynomials and overdetermined systems of PDE's. 

As the classification of superintegrable systems progressed, it was discovered how their symmetry algebra are not only Lie algebra but  finitely generated polynomial algebras \cite{das10,hoq15,mil13,chen19,liao18,kuru20}. The construction of these symmetry algebras is usually obtained via involving calculations relying on differential operators, a particularly difficult task in higher dimensions. However, once the symmetry algebra is obtained one can construct some classes of finite dimensional unitary representations and Casimir invariants to study the spectrum algebraically. However, the scope is limited as these algebraic structures are often only defined by the action on the physical wavefunctions. It is even known that how one even can describe these algebras in terms of different presentations (for example the equitable presentation \cite{vin15}), due to existence of larger set of linearly independent integrals. In some cases \cite{das09}, like in three dimensions, one even needs these linearly independent integrals to close the symmetry algebra in a finitely generated polynomial algebra. The fact that these algebraic structures are constructed only via explicit realisations, naturally induce a limitation to their complete classification and an understanding of their intrinsic properties. 

The connection between integrals of motion and certains type of Lie algebra was established in the 90's \cite{cal99}. The case of the superintegrable systems on the sphere was studied 
and connected with the $su(3)$ Lie algebra. Later on, using explicit realisations and irreducible unitary representations, a quadratic algebra where the structure constants depend on partial Casimir operators was obtained \cite{cal06,cal09}. The Hamiltonian corresponds to what is now refereed as the generic model on the 2-sphere. The significance of this generic two dimensional model was demonstrated \cite{post11,mil14} with the connection with orthogonal polynomials and the full Askey scheme. It was shown how all superintegrable systems on 2D conformally flat space, 59 divided in 12 classes are in fact related via 
an Inon\"u--Wigner contraction of their quadratic algebra. The symmetry algebra of this model and generalizations attracted attention recently \cite{kmp,gab18,vin19}. It was also connected to other types of algebraic structures such as the Bannai-Ito algebra. It was also discussed how the Racah algebra $R(3)$ can in fact be understood with the Howe duality \cite{vin19} in regard of the pair $(O(6), su(1,1))$.
\newline
\newline
In this paper, we will point out how without referring to an explicit realisation or representation of the underlying Lie algebra, we can build polynomials of degree 2, 3 and 4 in the enveloping algebra. They can be interpreted as integrals of motion. 
The polynomial algebras they generate are in fact of higher order and no longer of a simple quadratic form, but cubic or quartic. It is possible to express them in different ways which may involve, depending of the choice of basis, structure constants which are polynomials of the Cartan elements and Casimir invariants. A lower (quadratic) polynomial algebra arises when one refer to explicit coordinate representations acting on explicit statese, but it is then a consequence of further constraints among the integrals in the enveloping Lie algebras. Our observation also points out also that there is some freedom when one refers to a polynomial algebra of superintegrable systems: while some of them seems to be, after all, consequences of explicit choices of realization, here we propose other ones that can be defined only by the properties of Lie algebras, their MASA's and particular polynomials in their enveloping algebras. From this point of view, we also offer a way to classify such objects and probably allows to study a broader class of models by looking at other type of realisations.
\newline

The paper is organized in the following way. In section \ref{sec2}, we recall the quadratic Racah algebra for the classical and quantum superintegrable systems on the 2-sphere. The construction of the model on the 2-sphere is presented in Section \ref{sec3}, using the $su(3)$ algebra and  its corresponding quadratic integrals of motion. In Section \ref{sec4}, we demonstrate that with an appropriate combination of second order polynomials which commute with the Cartan elements, we can obtain a cubic algebra with structure constants given by integer numbers. We exploit as well the quadratic and cubic Casimir to present other form of the symmetry algebra. In particular, an algebra which involve polynomials of degree at most 2 for the element of the commutant and at most degree 4 in terms of the Cartan. We also show how the cubic algebra reduces to the second order one in the classical case. In Section \ref{sec5}, we present a contraction of the previous model, obtaining a modified six-dimensional cubic algebra. In the Appendix A, we show that the previous cubic algebra can also be written more generically as a quartic one.

\section{ Explicit model } \label{sec2}

The system we consider is the two dimensional superintegrable Hamiltonian $H$ associated with the $su(3)$ algebra \cite{cal06} living in the 2-sphere $s_1^2+s_2^2+s_3^2=1$, 
\begin{align}\label{hamil}
H=\frac{1}{2}\left(p_1^2+p_2^2+p_2^3+\frac{\alpha_1^2}{s_1^2}+\frac{\alpha_2^2}{s_2^2}+\frac{\alpha_3^2}{s_3^2}\right)=\frac{1}{2}\sum_{i<j}\left(s_i p_j -s_j p_i \right)^2 +\frac{1}{2}\sum_i \frac{\alpha_i}{s_i^2}\, ,
\end{align}
where $\alpha_i$ are real numbers. The spherical system (\ref{hamil}) is nothing else than an embedding of the unit sphere into the flat three dimensional space, considering also the constraint $s_1 p_1+s_2 p_2+s_3 p_3=0$. When we parametrize the 2-sphere with the spherical coordinates $s_1=\cos \theta \cos \phi$, $s_2=\cos \theta \sin \phi$ and $s_3=\sin \theta$, the above Hamiltonian takes the form 
\begin{align}
H= \frac{1}{2} \left(  p_{\theta}^{2}  + \frac{ p_{\phi}^{2}}{\cos^{2}\theta}+\frac{1}{\cos^{2}\theta}\left( \frac{\alpha_{1}^2}{\cos^{2}\phi} + \frac{\alpha_{2}^2}{\sin^{2}\phi}\right) + \frac{\alpha_{3}^2}{\sin^{2}\theta}  \right)\, ,  
\end{align}
where we can easily check that after separation of variables, appear the well-known one dimensional P\"oschl-Teller class of potentials. It is also known the fact that the Hamilton-Jacobi equation transform into a system of two ordinary differential equations, which can be solved analytically \cite{cal99}. As it is already known, both separation and solvability is related with the existence of certain integrals of motion. Indeed, the system displays three conserved quantities which may be written as\footnote{Our notation in the conserved quantitites slightly differs to the one in \cite{post11}, but there is no major changes.}
\begin{equation}\label{tint}
T_\ell=-\frac{1}{4}\left[ \left(s_j p_k -s_k p_j\right)^2+\left(\alpha_{j}\frac{s_{k}}{s_{j}}+ \alpha_{k}\frac{s_{j}}{s_{k}}\right)^2+(\alpha_j{-}\alpha_k)^2 \right], \quad j\neq k\neq \ell
\end{equation}
and using the sphere constraint the Hamiltonian (\ref{hamil}) is written
as a sum of such integrals of motion
\begin{equation}\label{newh}
H=-2 (T_1+T_2+T_3) -\sfrac{1}{2}(\alpha_1^2+\alpha_2^2+\alpha_3^2)\, ,
\end{equation}
where from now the constant additive term depending on the coupling constants will be omitted.
We can study the model and its symmetries from both classical and quantum approaches, where there are no big differences in the resulting structures. The idea is also to present how those realisations are in contrast with what we will obtain following the purely algebraic approach of Section \ref{sec3}.

At the classical level, in order to find the resulting Poisson bracket structure, we first note that the three integrals and the Hamiltonian satisfy
\begin{align}\label{claspoi}
\{ H, T_i\}_\text{\tiny {\rm PB}}=0,  \quad \{ T_i, T_j \}_\text{\tiny {\rm PB}}=T_{ij} \, ,
\end{align}
where we have denoted $T_{ij}$ as a new object, but as we shall prove below, it is not completely independent. In fact, this is already the case for some of the $T_{ij}$, where we have $T_{12}=-T_{13}=T_{23}$. If we restrict to only consider $T_{12}$, the further Poisson brackets give us
\begin{align} \label{tc121}
\{ T_{12} ,T_1\}_\text{\tiny {\rm PB}}&=2 T_1 T_3-2 T_1 T_2+\sfrac{1}{2}(\alpha_2^2{-}\alpha_3^2)(T_1+T_2+T_3+\sfrac{1}{4}[2\alpha_1^2{+}\alpha_2^2{+}\alpha_3^2]) \, 
\end{align}
where we can obtain the analogous relations $\{ T_{12} ,T_\ell\}_\text{\tiny {\rm PB}}$ for $\ell=2,3$ making cyclic permutations to the r.h.s of (\ref{tc121}). Clearly $\{ T_{12} ,T_1\}_\text{\tiny {\rm PB}}+\{ T_{12} ,T_2\}_\text{\tiny {\rm PB}}+\{ T_{12} ,T_3\}_\text{\tiny {\rm PB}}=0$. In this manner we find the system is maximally superintegrable choosing $H$ and the integrals $T_i$ with $i=1,2,3$. As a consequence, the superintegrablity show us that $T_{12}$ should not be an independent object and satisfies the following simple functional relation,
\begin{equation}
T_{12}^2=- 4T_1 T_2 T_3-\sum_\ell^3 T_\ell \sum_i^3 \alpha _i^2 T_i^2-\sfrac{1}{4}\sum_\ell^3 (\alpha_\ell^4{+}3\alpha_\ell^2(\alpha_j^2{+}\alpha_k^2){+}\alpha_j^2\alpha_k^2)T_\ell-\sfrac{1}{8}\prod_{m<n}^3 (\alpha_m^2+\alpha_n^2), \quad j\neq k\neq \ell
\end{equation}
\newline\newline
The quantization procedure of the system (\ref{hamil}) does not introduce radical changes of the symmetry structure discussed above. The resulting operator algebra is known as the Racah $R(3)$ algebra and is a direct consequence of the differential operator realisation. This kind of algebras have been extended to models on $n$ dimensional sphere \cite{vin19} and on the pseudo sphere \cite{kuru20}. In the quantum scenario, the integrals (\ref{tint}) can be rewritten as
\begin{equation}
\hat{T}_\ell=-\frac{1}{4}\left[ -\hbar^2 \left(s_j \partial_k -s_k \partial_j\right)^2+\left(\alpha_{j}\frac{s_{k}}{s_{j}}+ \alpha_{k}\frac{s_{j}}{s_{k}}\right)^2+(\alpha_j{-}\alpha_k)^2 -\hbar^2 \right], \quad j\neq k\neq \ell \, ,
\end{equation}
where, up to an additive constant term, we also have  $\hat{H}=-2 (\hat{T}_1+\hat{T}_2+\hat{T}_3)$. Following the same track as in the classical case in (\ref{claspoi}), we define new quantum operators $\hat{T}_{ij}$ from the algebra,
\begin{equation}
[\hat{H}, \hat{T}_i]=0, \quad \hat{T}_{ij}=\frac{1}{i\hbar} [\hat{T}_i, \hat{T}_j ] \, .
\end{equation}
The further commutation relations show us also the quantum clousure version of (\ref{tc121}) in terms of previously known integrals
\begin{equation}\label{t121quan}
\frac{1}{i \hbar}[\hat{T}_{12}, \hat{T}_1 ] =\{ \hat{T}_1, \hat{T}_3\}-\{\hat{T}_1,\hat{T}_2\}+\sfrac{1}{2}(\alpha_2^2{-}\alpha_3^2)(\hat{T}_1+\hat{T}_2+\hat{T}_3+\sfrac{1}{4}(2\alpha_1^2{+}\alpha_2^2{+}\alpha_3^2-3\hbar^2))\, ,
\end{equation}
satisfying also the relations $\hat{T}_{12}=-\hat{T}_{13}=\hat{T}_{23}$, which show us there is no further independent quantities. Indeed, as in the classical case, the square of the operators $\hat{T}_{12}$ can be expressed as,
\begin{align}
\hat{T}_{12}^2&=-4 \{ \hat{T}_{1}{+}\sfrac{2}{3}\hbar^2, \hat{T}_{2}{+}\sfrac{2}{3}\hbar^2, \hat{T}_{3}{+}\sfrac{2}{3}\hbar^2 \}-\{ \textstyle \sum_i \hat{T}_{i}, \textstyle  \sum_\ell(\alpha_\ell^2{-}\sfrac{3}{8}\hbar^2)\hat{T}_{\ell} \}+ \\ \notag
&-\sfrac{1}{4}\textstyle \sum_\ell^3 (\alpha_\ell^4{+}3\alpha_\ell^2(\alpha_j^2{+}\alpha_k^2){+}\alpha_j^2\alpha_k^2 -\sfrac{1}{18} \hbar^2 [60 \alpha _\ell^2{+}6 \alpha _j^2{+}6 \alpha _k^2{+}155 \hbar^2] )T_\ell-\sfrac{1}{8}\textstyle\prod_{m<n}^3 (\alpha_m^2{+}\alpha_n^2)+\\ \notag
&+\sfrac{\hbar^2}{48}(\textstyle \sum_\ell \alpha_\ell^2(\alpha_\ell^2{-}3\hbar){+}19\sum_{i<j} \alpha_i^2 \alpha_j^2)+\sfrac{1805}{1728} \hbar^6 \, .
\end{align}
It is straightforward to verify that in the limit $\hbar \rightarrow 0$ all these expressions reproduce the classical results discussed above. It is interesting to remark that in order to understand the origin and nature of the symmetry algebra, the intertwining operators approach  \cite{cal06} display a key role when they are identified with the elements of the algebra.

\section{Algebraic integrability from $su(3)$ }\label{sec3}

The superintegrable systems presented in Section \ref{sec3} can be obtained from a $su(3)$ Lie algebra \cite{cal99,cal06,cal09}. This algebra was constructed using the knowledge of three types of intertwining operators, which act on the wavefunctions which correspond to irreducible unitary representations (IUR). We use the following $su(3)$ basis   
\begin{equation}\label{xxalgebra}
\begin{array}{llllll}
 [X_1,X_1]=0, && [X_1,X_2]=0, && [X_1,X_3]=2 X_4 \\ \null
  [X_1,X_4]=-2 X_3,   && [X_1,X_5]=X_6,    &&     [X_1,X_6]=-X_5  \\ \null
 [X_1,X_7]=-X_8,&& [X_1,X_8]=X_7,&& [X_2,X_2]=0 \\  \null
 [X_2,X_3]=-X_4,&& [X_2,X_4]=X_3,&& [X_2,X_5]=X_6 \\  \null
 [X_2,X_6]=-X_5,&& [X_2,X_7]=2 X_8,&& [X_2,X_8]=-2 X_7 \\  \null
 [X_3,X_3]=0,&& [X_3,X_4]=2 X_1,&& [X_3,X_5]=-X_7 \\ \null
 [X_3,X_6]=-X_8,&& [X_3,X_7]=X_5,&& [X_3,X_8]=X_6 \\  \null
 [X_4,X_4]=0,&& [X_4,X_5]=X_8,&& [X_4,X_6]=-X_7 \\  \null
 [X_4,X_7]=X_6,&& [X_4,X_8]=-X_5,&& [X_5,X_5]=0 \\  \null
 [X_5,X_6]=2 X_1+2 X_2,&& [X_5,X_7]=-X_3,&& [X_5,X_8]=X_4 \\ \null
 [X_6,X_6]=0,&& [X_6,X_7]=-X_4,&& [X_6,X_8]=-X_3   \\  \null
  [X_7,X_7]=0,&& [X_7,X_8]=2 X_2,&& [X_8,X_8]=0 
 \end{array}
\end{equation}
to study the properties of the related symmetry algebra of the model in section \ref{sec2}. There are other choices of basis, for instance the one introduced in \cite{cal09},
\begin{align}\label{aope}
A^{\pm}&=\sfrac{1}{2}(\pm  X_3 - i X_4 ), \quad\quad A=-\sfrac{1}{2} i X_1 \\\label{bope}
B^{\pm}&=\sfrac{1}{2}(\pm X_5 - i X_6), \quad\quad B=A+C \\
C^{\pm}&=\sfrac{1}{2}(\pm X_7 - i X_8), \quad\quad C=-\sfrac{1}{2} i X_2 \label{cope}
\end{align}
Each of the pairs $A^{\pm}$, $B^{\pm}$ and $C^{\pm}$ were studied from their action on the physical states of the quantum model introduced in Section \ref{sec2}. All the states 
connected by intertwining operators (\ref{aope}), (\ref{bope}) and (\ref{cope}) can be written in terms of orthogonal polynomials, and they are associated to irreducible unitary representations of $su(3)$. 
In this way, the states linked by the action of the operators (\ref{aope})-(\ref{cope}) can be interpreted as belonging to physical states of different Hamiltonians characterized by different
eigenvalues of the Cartan generators $X_1, X_2$. This is why those intertwining operators turns out to be related with the ones in the shape invariant potentials and they also can be used to construct a generating spectrum algebra. The next consequence is that the products of two adjoint related intertwining operators such ass $A^{+}A^{-}$, $B^{+}B^{-}$ and $C^{+}C^{-}$ are algebraic integrals, in the same spirit as in supersymmetric quantum mechanics \cite{coo95} or also more recently with intertwining operators in the Calogero models \cite{cor14}. A generating spectrum algebra which is valid only on the IUR of the underlying lattice of Hamiltonians was obtained. 

Casimir operators are well known \cite{pop68}  for all simple Lie algebras and they are play an important role in the context of symmetries in quantum mechanics. From the set generators of $su(3)$, $\{X_{1},X_{2},X_{3},X_{4},X_{5},X_{6},X_{7},X_{8}\}$, it is possible to build a second and third order Casimir invariants,
\begin{align}
{\cal C}_{2}&= -\sfrac{2}{3} (X_1^{2}+X_2 X_1+ X_{2}^{2}) -\sfrac{1}{2} (X_3^{2} + X_4^{2} + X_5^{2} +X_6^{2} +X_7^{2}+ X_8^{2})\, , \\ \label{cas3}
{\cal C}_{3}&=(X_8 X_6{+}X_7 X_5) X_4{+}(X_8 X_5{-} X_7 X_6) X_3{+}\sfrac{4}{27}\left(X_1{-}X_2\right)\left(2X_1{+}X_2\right)\left(X_1{+}2X_2\right)+\\ \notag
&+\sfrac{1}{6}\{X_1{+}2X_2,X_3^2{+}X_4^2\}{+}\sfrac{1}{6}\{X_1{-}X_2,X_5^2{+}X_6^2\}{-}\sfrac{1}{6}\{2X_1{+}X_2,X_7^2{+}X_8^2\}{-}\sfrac{4}{3}(X_1{-}X_2)\, .
\end{align}
Both Casimir invariants hide some information of the algebra structure and also, as we shall see, of the polynomial algebra. Indeed, in this construction the Hamiltonian and the integrals are generated via polynomials in the enveloping algebra of $su(3)$, where $X_{1}$ and $X_{2}$ are associated with the Cartan generators (MASA) \cite{cal06}. The set of commuting operators or integrals of motion can be obtained as 
\begin{align}
T_{1}&=-\sfrac{1}{4}\left(X_{7}^{2}+X_{8}^{2}+X_{2}^2 \right) \, ,\label{t1m}\\
T_{2}&=-\sfrac{1}{4}\left(X_{5}^{2}+X_{6}^{2} +X_{1}^2+2X_1X_2+X_{2}^2\right)\, ,  \label{t2m} \\
T_{3}&=-\sfrac{1}{4}\left(X_{3}^{2}+X_{4}^{2}+X_{1}^2\right)\, , \label{t3m}  
\end{align}
where in comparison with the ones constructed in \cite{cal06}, we have included some extra terms depending on the Cartan generators whose relevance will be understood below.
The second order Casimir operator can be expressed in terms of the $T_i$ operators,
\begin{equation}
{\cal C}_2=2\left(T_1+T_2+T_3\right) +\sfrac{1}{3}(X_1^2+X_1X_2+X_2^2)\label{C2}
\end{equation}
and therefore the Hamiltonian  as a function ${\cal C}_2$,
\begin{equation}
 H=-2(T_{1} + T_{2} + T_{3})= -{\cal C}_{2} +\sfrac{1}{3}(X_1^2+X_1X_2+X_2^2) \, .\label{H}
\end{equation}
In the above expression, the comparison with the Hamiltonian in (\ref{hamil}) or (\ref{newh}) is valid up to an additive constant. Such a freedom is related with the fact that any combination of the Cartan generators and Casimir operators can be chosen as the Hamiltonian. In particular, in the quantum case correspond to the physical expression when using the explicit differential operator realisation \cite{cal99}. In the other basis, we can express the integrals $T_i$ in terms of $A$, $A^{\pm}$, $B$, $B^{\pm}$, $C$ and $C^{\pm}$ in the following way,
\begin{align}
T_{1}&=\sfrac{1}{2}\{C^+,C^- \}+C^2 \, ,  \label{t1c}\\
T_{2}&=\sfrac{1}{2}\{B^+,B^- \}+B^2\, , \ \label{t2c} \\
T_{3}&=\sfrac{1}{2}\{A^+,A^- \}+A^2\ , \ \label{t3c}  
\end{align}
The Hamiltonian (\ref{H}) in this basis has the simple form
\begin{equation}
H=-(A^{+}A^{-}{+} B^{+}B^{-}{+}C^{+}C^{-})- 2(A^2{+}B^2{+}C^2)\, ,
\end{equation}
and the following relations hold,
\begin{align}
A^{+} H&= (H{+} 2A{-}1)A^{+} \, , \\
A^{-} H&= (H{-} 2A{-}1)A^{-} \, , \\
B^{+} H&= (H{+} 2B{-}1)B^{+}\, , \\
B^{-} H&= (H{-} 2B{-}1)B^{-}\, , \\
C^{+} H&= (H{-}2C{-}1)C^{+}\, , \\
C^{-} H&= (H{-} 2C{-}1)C^{-} \, .
\end{align}
where we have also
\begin{equation}
[C^\pm,T_{1}]=0, \quad [B^\pm,T_{2}]=0, \quad [A^\pm,T_{3}]=0 \, .
\end{equation}
It is worth to remark here that these are purely algebraic relations, without specifying any realisation, and appears in a similar fashion as the intertwining relations \cite{coo95} in supersymmetric quantum mechanics or related with Darboux transformations. However as we deal here with a higher dimensional cases and considering they are related to several conserved quantities, they look like intertwining operators for the Calogero models 
\cite{cor14}. We focus now into the construction of a cubic and higher order algebras based on the polynomials elements of $su(3)$.

\section{Further elements of the enveloping algebra and polynomial algebra}\label{sec4}
In order to describe the polynomial cubic algebra, we start building further elements via commutation relations, 
\begin{align}
T_{12}&= [ T_{1},T_{2}]\, , \label{t12} \\ 
T_{13}&= [ T_{1},T_{3}] \, ,\label{t13}  \\  
T_{23}&= [ T_{2},T_{3}]\, , \label{t23}
\end{align}
but, in the same footing as the classical and quantum schemes, these new integrals are not all independent: $T_{13}=-T_{12}$ and $ T_{23}=T_{12}$. These polynomials are part of the commutant in regard to the Cartan generators, i.e., they satisfy $[T_{12},X_{i}]=[T_{13},X_{i}]=[T_{23},X_{i}]=0$ for $i=1,2$. As only one of the three polynomials of degree 3 are independent, we are free to choose
\begin{align}
T_{12} &=\sfrac{1}{4} (X_8 X_6 X_3{-}X_8 X_5 X_4{+}X_7 X_6 X_4{+}X_7 X_5 X_3{-}X_8^2{-}X_7^2{+}X_6^2{+}X_5^2{-}X_4^2{-}X_3^2) \\
&=\sfrac{1}{4} \{ X_8, X_6, X_3 \}{-}\sfrac{1}{4}\{ X_8, X_5, X_4\}{+}\sfrac{1}{4}\{X_7, X_6, X_4\}{+}\sfrac{1}{4}\{X_7, X_5, X_3\}
\end{align}
where $\{a,b,c\}=\sfrac{1}{6}(abc+cab+bca+acb+bac+cba)$. Next, it will generate the following polynomials in the enveloping algebras
\begin{align}
T_{121}&= [ T_{12},T_{1}]\, , \label{t121} \\   
T_{122}&= [ T_{12},T_{2}]\, , \label{t122} \\ 
T_{123}&= [ T_{12},T_{3}]\, . \label{t123}  
\end{align}
Again, we have that $[ T_{121},X_{i}]=[T_{122},X_{i}]=[T_{123},X_{i}]=0$ for $i=1,2$ and they can be seen as fourth degree polynomials 
with the explicit forms,
\begin{align} \label{t121}
 T_{121}&= \sfrac{1}{16} \{X_4^2{+}X_3^2{-}X_6^2{-}X_5^2,X_8^2{+}X_7^2\}{-}\sfrac{1}{8} 
\{  (X_8 X_6{+}X_7 X_5) X_4{+}(X_8 X_5{-} X_7 X_6) X_3, X_2\}{-}\sfrac{1}{2}X_2^2 \\ 
 T_{123}&= \sfrac{1}{16} \{X_6^2{+}X_5^2{-}X_8^2{-}X_7^2,X_4^2{+}X_3^2\}{-}\sfrac{1}{8} 
\{  (X_8 X_6{+}X_7 X_5) X_4{+}(X_8 X_5{-} X_7 X_6) X_3, X_1\}{+}\sfrac{1}{2}X_1^2  \\ \label{laseq}
 T_{122}&=-T_{121}-T_{123}\,.
\end{align}
These four degree elements cannot be expressed as polynomials in terms of the algebra elements $T_i$ or $T_{ij}$, so we can treat them as new elements in the algebra. Although, by comparison with the Eq. (\ref{cas3}) it is possible to express such new integrals in terms  of the the cubic Casimir invariant, we will return to this point later in Section \ref{4p1}. The relation (\ref{laseq}) tell us that only two of the new objects are independent and there is again a freedom to choose the basis for the fourth degree polynomials. We choose then the complete basis \{$T_{1}$, $T_{2}$, $T_{3}$, $T_{12}$, $T_{121}$, $T_{122}$, $X_{1}$, $X_{2}$ \} and show that they close in a polynomial algebra, i.e. taking consecutive commutators one can write the results in terms of a polynomial combination of the already defined polynomials: 
\begin{align}
[T_{121},T_1]&=2\{ T_{12},T_1 \} \\
[T_{122},T_2]&=2\{ T_{12},T_2 \} \\
[T_{121},T_2]&=[T_{122},T_1]=-\{ T_{12},T_1 \}-\{ T_{12},T_2 \}+\{ T_{12},T_3 \} \\
[T_{121},T_3]&=-\{ T_{12},T_1 \}+\{ T_{12},T_2 \}-\{ T_{12},T_3 \} \\
[T_{122},T_3]&=\{ T_{12},T_1 \}-\{ T_{12},T_2 \}-\{ T_{12},T_3 \}  
\end{align}  
\begin{align} 
[T_{121},T_{12}]&=2\{ T_{122},T_1 \} +\{ T_{121},T_1 \}+\{ T_{121},T_2 \}-\{ T_{121},T_3 \}\\
[T_{122},T_{12}]&=-2\{ T_{121},T_2 \} -\{ T_{122},T_1 \}-\{ T_{122},T_2 \}+\{ T_{122},T_3 \} \end{align}
\begin{align} \notag
[T_{121},T_{122}]&= -\{\{ T_{12}, T_1\}, T_1\}{-}\{ \{ T_{12}, T_2\}, T_2\}{-}\{ \{ T_{12}, T_3\}, T_3\}{+}2\{ \{ T_{12}, T_1\}, T_2\}\\
&{+}2\{ \{ T_{12}, T_1\}, T_3\}{+}2\{ \{ T_{12}, T_2\}, T_3\}
\end{align}

From the last relation the cubic nature of the polynomial algebra is verified. Remark also that the Cartan generators do not appear explicitly in the above algebra. However, this result is kind of particular and is indeed the lowest degree of the algebra we can obtain by taking further nested commutators (i.e. $T_{i_{1},...,i_{k}}=[...[T_{i_{1}},T_{i_{2}}],....,T_{i_{k}}]$ ) of  elements of the commutant. We define the degree of the algebra by the highest degree of the polynomials on right side of the commutation relations in terms of both the commutant and the Cartan generators.  This is related with the appropriate combination of the Cartan generators chosen to define $T_i$ in (\ref{t1m}),  (\ref{t2m}) and  (\ref{t3m}). It is possible, for instance, to remove the Cartan generators in those equations, yielding $T_{1}=-\sfrac{1}{4}\left(X_{7}^{2}+X_{8}^{2}\right) $, $T_{2}=-\sfrac{1}{4}\left(X_{5}^{2}+X_{6}^{2}\right)$, $T_{3}=-\sfrac{1}{4}\left(X_{3}^{2}+X_{4}^{2}\right)$. In this case the resulting polynomial algebra is quartic an depends explicitly on the Cartan generators. The full algebra is described in the Appendix \ref{A1}. We can alternatively rewrite the algebras directly in terms of the quadratic or cubic Casimir invariants. This fact points out how using some explicit realisations one can obtain further constraints on the Casimir invariants, for example reducing the algebra to certain sectors. There exists another way to express the algebra using the elements (\ref{t1c}),  (\ref{t2c}) and  (\ref{t3c}) in terms of the operators $A$'s, $B$'s and $C$'s. When we calculate the operator $T_{12}$ in (\ref{t12}), we can denote can be written as
\begin{align}
[T_1,T_2]=-Y_1+Y_2
\end{align}
where
\begin{equation}
Y_1=A^{+} C^{+} B^{-}\quad\quad \text{and}\quad  \quad Y_2= Y_1^\dagger=B^{+} C^{-} A^{-} \, .
\end{equation}
In this new basis, we can choose the set  $\{A,B,T_1,T_2,T_3,Y_1,Y_2\}$ and check that the algebra can be closed in analogue form as \cite{cal09}, where in particular we have
\begin{align}
[Y_1,Y_2]&=2(A{-}1)(T_1T_2{+}Y_1{+}Y_2){-}2(B{-}1)(T_1T_3{+}2Y_2){+}2(B{-}A)T_2T_3{-}2AB(B{-}A)(T_1{+}T_2{+}T_3)\\ \notag
&+2AB(B{-}A)(A^2{-}AB{+}B^2{-}3){+}2(B^2{-}A^2)(T_1{-}(B{-}A)^2{+}1){+}2(B{-}A)^2 (T_2{-}T_3){-}2(B{-}A)(T_1{-}T_2{-}T_3) \, .
\end{align}
In terms of this basis the algebra is of fifth order,  which will be relevant also to understand the reduction for the classical and quantum cases. In what follows, we demonstrate how such algebras are reduced to quadratic order. Notice that the three degree elements $Y_1,Y_2$
are related by means of the cubic Casimir, 
\begin{align}
Y_1{+ }Y_2 {-}\sfrac{1}{4}{\cal C}_3&=\sfrac{2}{3}(B{-}A)(T_1{+}T_2{-}2T_3){+}\sfrac{2}{3}A(2T_1{-}T_2{-}T_3){-}T_1{+}T_2{+}T_3\\
&{+}\sfrac{2}{3}(A{+}B){+}2AB{+}\sfrac{10}{27}(2A{-}B)(A{+}B)(A{-}2B)
\end{align}

\subsection{Reduction of the higher order polynomial algebra to quadratic ones}\label{4p1}

The cubic polynomial algebra discussed above is inherent of the algebraic $su(3)$ structure coming from (\ref{xxalgebra}).  In the classical and quantum schemes, they get reduced into a second order one.  In order to understand in details how this truncation occurs in classical mechanics we are going to study separately the two pictures under different points of view. In the classical scheme we used the following realization in terms of coordinates and momenta, following the prescription in \cite{cal99},
\begin{equation}\label{classi}
\begin{array}{llllll}
X_1=\alpha_2-\alpha_1&& X_2=\alpha_3-\alpha_2 \\ [2ex] 
X_3=x_1p_2 -x_2p_1    && X_4=-\alpha_2\displaystyle \frac{x_1}{x_2}-\alpha_1\displaystyle \frac{x_2}{x_1}   \\ [2.1ex]
X_5=x_1p_3 -x_3p_1    && X_6=-\alpha_1 \displaystyle\frac{x_3}{x_1}-\alpha_3\displaystyle \frac{x_1}{x_3},   \\ [2.1ex]
X_7=x_2p_3 -x_3p_2    && X_8=-\alpha_3\displaystyle \frac{x_2}{x_3}-\alpha_2 \displaystyle\frac{x_3}{x_2}.  
 \end{array}
\end{equation}
Where we can see that Cartan generators $X_1$ and $X_2$ in this realisation depend only on the coupling constants, so they are trivially 
conserved quantities. Plugging in this realisation in the definition of the integrals (\ref{t1m}),  (\ref{t2m}) and  (\ref{t3m}), we exactly reproduces (\ref{tint}) and the Poisson brackets algebra in Section \ref{sec1}. In particular, the relation (\ref{t121}), can be rewritten in terms of the operators $T_i$,
\begin{align} \notag
 T_{121}&= \{T_3,T_1\}{-} \{T_2,T_1\}{-}\sfrac{1}{2}(2X_1X_2{+}X_2^2)T_1{+}\sfrac{1}{2}X_2^2(T_3-T_2){-}\sfrac{1}{8} 
\{  (X_8 X_6{+}X_7 X_5) X_4{+}(X_8 X_5{-} X_7 X_6) X_3, X_2\}\\
&-\sfrac{1}{2}X_2^2{-}\sfrac{1}{8}(2X_1X_2+X_2^2)X_2^2 \,,
\end{align}
where we can identify the leading term when is compared with Eqs. (\ref{tc121}) and (\ref{t121quan}). Using the fact the definitions of $X_1$ and $X_2$ and 
\begin{align}
(X_8 X_6{+}X_7 X_5)X_4{+} (X_8 X_5{-} X_7 X_6) X_3=4(\alpha_1 T_1{+}\alpha_2 T_2{+}\alpha_3 T_3){+}\alpha_1(\alpha_2{-}\alpha_3)^2{+}(\alpha_2{+}\alpha_3)(\alpha_1^2{+}\alpha_2 \alpha_3)
\end{align}
and ommiting the constant term proportional to $X_2^2$, the  $T_{121}=\{ T_{12} ,T_1\}_\text{\tiny {\rm PB}}$ takes the form
 \begin{align} \notag
\{ T_{12} ,T_1\}_\text{\tiny {\rm PB}}=&2T_3T_1{-}2T_2T_1{-}\sfrac{1}{2}(2 \alpha_1{-}\alpha_2{-}\alpha_3)(\alpha_2{-}\alpha_3)T_1{-}\sfrac{1}{2}(\alpha_2{-}\alpha_3)^2(T_2{-}T_3)+(\alpha_2{-}\alpha_3)(\alpha_1 T_1{+}\alpha_2 T_2{+}\alpha_3 T_3)\\
&+\sfrac{1}{4}\alpha_1(\alpha_2{-}\alpha_3)^3{+}\sfrac{1}{4}(\alpha_2{-}\alpha_3)^2(\alpha_1^2{+}\alpha_2\alpha_3)+\sfrac{1}{8}(\alpha_2-\alpha_3)^3(-2\alpha_1+\alpha_2+\alpha_3)
\end{align}
Then, by a simple manipulation we can easily prove the expression reduces to the form of the Poisson bracket (\ref{tc121}), and therefore the cubic polynomial algebra truncates into a second order one. \newline

In order to understand a different point of view of the algebra truncation, we again rewrite down the relation (\ref{t121}) in terms of the cubic Casimir (\ref{cas3})
\begin{equation}
 T_{121}=[T_{12},T_{1}]= \{T_3,T_1\}{-} \{T_2,T_1\}{-}\sfrac{1}{4}X_2 {\cal C}_3{-}\sfrac{1}{12}X_2(2X_1{+}X_2){\cal C}_2{-}\sfrac{1}{216}X_2(X_1{+}X_2)(36{+}(2X_1{+}X_2)^2)\, .
\end{equation}
which can be seen as a quadratic relation from the cubic Casimir which is as well purely algebraic,  where the structure constants depend as higher order polynomials on Cartan generators.The other relation is given by the following expression
\begin{equation}
 T_{122}=[T_{12},T_{2}]= \{T_1,T_2\}{-} \{T_2,T_3\}{+}\sfrac{1}{4}(X_1{+}X_2) {\cal C}_3{-}\sfrac{1}{12}(X_1^2{-}X_2^2){\cal C}_2{-}\sfrac{1}{216}(X_1^2{-}X_2^2)(36{+}(X_1{-}X_2)^2)\, .
\end{equation}
As we pointed out before, polynomial algebras of order higher than three can be obtained using different basis or elements of the algebra. In this case, we can clearly see how when the Cartan generators takes constant values from (\ref{classi}), the expression reduces to
\begin{align}\notag
 T_{121}&= \{T_3,T_1\}{-} \{T_2,T_1\}{-}\sfrac{1}{4}(\alpha_3{-}\alpha_2){\cal C}_3{+}\sfrac{1}{13}(\alpha_3{-}\alpha_2)(\alpha _1{+}\alpha _2{-}2 \alpha _3){\cal C}_2\\
 &{-}\sfrac{1}{216}(\alpha_3{-}\alpha_2)(\alpha _1{+}\alpha _2{-}2 \alpha _3)(36{+}(\alpha _1{+}\alpha _2{-}2 \alpha _3)^2) \, .
\end{align}
By virtue of the relation (\ref{C2}) we can remove any of the generators $T_i$ and inspect the resulting algebra. Without any loss of generality, we can take out $T_3$  consider the algebra generated by $\{T_1,T_2,T_{12}\}$ the Cartan generators and the second and third order Casimir. The resulting algebra takes the form,
\begin{align}\notag
[T_{12},T_{1}]&= -2T_1^2{-}2\{T_2,T_1\}{-}\sfrac{1}{4}X_2 {\cal C}_3{+}\left({\cal C}_2-\sfrac{1}{3}[X_1^2{+}X_1X_2{+}X_2^2]\right)T_1{-}\sfrac{1}{12}X_2(2X_1{+}X_2){\cal C}_2
 \\ \label{alg1} &-\sfrac{1}{216}X_2(X_1{+}X_2)(36{+}(2X_1{+}X_2)^2)\, , \\\notag
[T_{12},T_{2}]&=2T_2^2+ 2\{T_1,T_2\}{+}\sfrac{1}{4}(X_1{+}X_2) {\cal C}_3{-}\left({\cal C}_2-\sfrac{1}{3}[X_1^2{+}X_1X_2{+}X_2^2]\right)T_2{-}\sfrac{1}{12}(X_1^2{-}X_2^2){\cal C}_2\\ \label{alg2}
&-\sfrac{1}{216}(X_1^2{-}X_2^2)(36{+}(X_1{-}X_2)^2)\, .
\end{align}
Taking into account these relations and their elements,  we show the Casimir operator of this algebra,
\begin{align}\notag
 K &= T_{12}^2{-}2 \{T_1^2, T_2\}{-}2\{T_2^2, T_1\}{+}4(T_1^2{+}T_2^2){+}(4{+}{\cal C}_2{-}\sfrac{1}{3}[X_1^2{+}X_1X_2{+}X_3])\{T_1, T_2\}+\\  \notag
 &-(2{\cal C}_2{+}\sfrac{1}{2}(X_1{+}X_2){\cal C}_3{-}\sfrac{1}{3}(3X_1^2{+}2X_1X_2{+}X_2^2){-}\sfrac{1}{6}(X_1^2{-}X_2^2){\cal C}_2{-}\sfrac{1}{108}(X_1{+}X_2)(X_1{-}X_2)^3)(T_1+1)\\ \label{Kcas}
 & -(2{\cal C}_2{+}\sfrac{1}{2}X_2{\cal C}_3{-}\sfrac{1}{3}(2X_1^2{+}X_2^2){+}\sfrac{1}{6}X_2(2X_1{+}X_2){\cal C}_2{+}\sfrac{1}{108}X_2(2X_1{+}X_2)^3)(T_2-1)\, .
\end{align}
The Casimir $K$  can be also be written in terms of the central elements only
\begin{align}
 K &=-\sfrac{1}{16}C_3^2{-}\sfrac{1}{36} C_2^2 \left([X_1{+}2 X_2]^2{+}9\right){-}\sfrac{1}{12} C_3 C_2 \left(X_1{+}2 X_2\right)\\ \notag
& -\sfrac{1}{216} C_3 \left(72 \left(2 X_1{+}X_2\right){+}\left(X_1{+}2 X_2\right) \left(4 X_1^2{-}11 X_2 X_1{-}11 X_2^2\right)\right) \\ \notag
 &-\sfrac{1}{648} C_2 \left(X_1{-}X_2\right) \left(\left(2 X_1{+}X_2\right) \left(4 X_1^2{+}7 X_2 X_1{+}7 X_2^2\right){-}36 \left(4 X_1{+}5 X_2\right)\right)\\\notag
&-\sfrac{1}{5832}\left([X_1{-}X_2]{}^2{+}36\right) \left(X_1{-}X_2\right) \left([2 X_1{+}X_2]{}^3{-}36 [X_1{+}2 X_2]\right).
\end{align}
\section{Contraction and Smorodinsky-Winternitz systems}\label{sec5}
It is known how from the $su(3)$ related differential operator realisation, one can obtain another quantum model via a contraction approach. In our case we follow this mechanism by setting
\begin{align*}
X'_{3}&= \frac{1}{R} X_{3}, \quad X'_{4}= \frac{1}{R} X_{4}\,  ,\\
X'_{7}&= \frac{1}{R} X_{7} , \quad X'_{8}= \frac{1}{R} X_{8}\, ,
\end{align*}  
and taking the singular limit
$R \rightarrow \infty$, the commutation relations in (\ref{xxalgebra}) yields
\begin{align*}
& [X'_{3},X'_{4}]=0,\quad [X'_{3},X'_{7}]=0, \quad [X'_{3},X'_{8}]=0\, ,\\
& [X'_{4},X'_{7}]=0,\quad [X'_{4},X'_{8}]=0,\quad [X'_{7},X'_{8}]=0\, .
\end{align*}
One can interpret these are related to Euclidean subalgebras which from a perspective of an underlying differential operator realisation to coordinates on an Euclidean space. The structure of the integrals is a similar polynomial (even if the element forming this polynomial satisfies a rather different algebra than $su(3)$).  However, now we have $T_{13}=0$ and $T_{12}-T_{13}=0$
\begin{align*}
 T_{12}&= \sfrac{1}{4} (X'^2_3{+}X'^2_4{+}X'^2_7{+}X'^2_8 {+}X'_3 X'_5 X'_7{+}X'_3 X'_6 X'_8{-}X'_4 X'_5 X'_8{+}X'_4 X'_6 X'_7) 
\end{align*}
The second order Casimir also changes in the contraction and now takes the form
\begin{align*}
 C_{2}&=\frac{1}{2} ( X'^2_{3} +X'^2_{4}+ X'^2_{7} +X'^2_{8} )\\
&= -2 T_1 -2 T_3  -\frac{1}{2}X'^2_1 -\frac{1}{2} X'^2_2 
\end{align*}
and thus would be provide another Hamiltonian, which may as well be expressed according to a physical space and differential operators in quantum mechanics or natural Hamiltonian in classical mechanics. Here we present the a entirely algebraic symmetry algebra. We can obtain analog expression for $T_{121}$, $T_{122}$, $T_{123}$ that we are omitting (again only two are independent as $T_{121}+T_{123}=0$).  We have the further relations, which form a six-dimensional cubic algebra
\begin{align*}
 [T_{121},T_1]&=0 \\
 [T_{122},T_{2}]&=2 \{T_{12},T_{2}\} \\
 [T_{121},T_{2}]&=- \{T_{12},T_{1}\} + \{T_{12},T_{3}\}+\sfrac{1}{2}(X_{1}^{2}-X_{2}^{2})T_{12}\\
 [T_{122},T_{1}]&=- \{T_{12},T_{1}\} + \{T_{12},T_{3}\}+\sfrac{1}{2}(X_{1}^{2}-X_{2}^{2})T_{12}\\
 [T_{121},T_{3}]&=0\\
 [T_{122},T_{3}]&=\{T_{12},T_{1}\}- \{T_{12},T_{3}\}-\sfrac{1}{2}(X_{1}^{2}-X_{2}^{2})T_{12}\\
 [T_{121},T_{12}]&= \{T_{121},T_{1}\} -\{T_{121},T_{3}\}-\sfrac{1}{2}(X_{1}^{2}-X_{2}^{2})T_{121}\\
 [T_{122},T_{12}]&= -\{T_{122},T_{1}\}-2 \{T_{121},T_{2}\}+ \{T_{122},T_{3}\}+\sfrac{1}{2}(X_{1}^{2}-X_{2}^{2})T_{122}\\
[T_{121},T_{122}]&=-\{\{T_{12},T_1\},T_1\} -\{\{T_{12},T_3\},T_3\}+ 2\{T_{121},T_{12}\}\\
&-X_{2}^{2} \{T_{12},T_{1}\}-X_{1}^{2} \{T_{12},T_{3}\}-\sfrac{1}{4}(X_{1}^{4}+X_{2}^{4})T_{12}
\end{align*}

In summary, we have shown that the contractions of the polynomial algebras
of symmetries can be easily implemented inside the formalism of polynomial algebras
in the universal enveloping algebras (of $su(3)$, in this case) by extending the well known
method of Inon\"u--Wigner. At the end we arrive to another polynomial algebra in the universal
covering algebra of the
contracted Lie group (in this case $isu(2)$).

\section{Conclusion}\label{sec6}
In this paper, we developed a new approach to the symmetry algebra of superintegrable systems. We introduced what can be seen as a six-dimensional quintic algebra where the Casimir and Cartan generator are part of the structure constant for $su(3)$. The algebra can be simplified in a six-dimensional cubic algebra with only integers as structure constants. We reformulated this results in a basis of $su(3)$ related to a generic two dimensional superintegrable systems on the 2-sphere. This algebra is in fact also related to underlying intertwining operators. We also reformulated this algebra using the quadratic and cubic Casimir invariants of $su(3)$ for which the algebra depend in at most degree 2 in the element of the commutant but still involve polynomials up to degree 4 in the Cartan generators. All these different algebraic structures do not need any explicit differential operator realizations and are valid beyond their action on physical wavefunctions. These results allow to provide algebraic characterization of the notion of superintegrability and their symmetry algebra. They allow to formulate what we refer as ``purely algebraic'' polynomial algebras. These results point out how the enveloping algebra of a Lie algebra can be used as the right underlying structure to formulate the symmetry algebras of superintegrable systems. 

These results also allow to show how classification and study can be made for certain classes of polynomial algebras based on further studies of enveloping Lie algebras. This also point out that such formulation of polynomial algebras may be relevant to study the algebraic properties of superintegrable systems, independently of any realization. 
The scope of the deformed oscillator algebra approach is still limited and had not been extended in context of $n$-dimensional systems. Only some partial results relying on substructure have
been exploited to present algebraic description of the spectrum. 
Algebraic constructions as the one of this paper offer alternative way to construct representations of polynomial algebras based on the representation theory of Lie algebras. Another application of this algebraic approach is the possibility on using new types of explicit realisations which would provide other examples of superintegrable models.

Other recent works on enveloping algebra \cite{cra20} of simple Lie algebras have been done in regard of labeling problem and Clebsch-Gordan coefficients \cite{cam19}. This point out how these approaches can have wider applicability.


\section*{Acknowledgement}
FC was partially supported by Fondecyt grant 1171475 and Becas Santander Iberoam\'erica. FC would like to thank the Departamento de F\'isica Te\'orica, \'Atomica y \'Optica at the Universidad de Valladolid and the University of Queensland for all the support and kind hospitality. IM was supported by by Australian Research Council Future Fellowship FT180100099. The authors would like to thank to the 
Centro de Ciencias de Benasque Pedro Pascual.  JN was partially supported by Junta de Castilla y Le\'on, Spain (BU229P18, VA137G18). 
\section{Appendix A: Quartic algebra relations}\label{A1}
We present here an alternative presentation of the algebra from the integrals of motion (\ref{t1m}),  (\ref{t2m}) and  (\ref{t3m}), but here eliminating the dependence of the Cartan generators,
\begin{align}
T_{1}&=-\sfrac{1}{4}\left(X_{7}^{2}+X_{8}^{2}\right) \label{t1} \\
T_{2}&=-\sfrac{1}{4}\left(X_{5}^{2}+X_{6}^{2}\right) \label{t2}  \\
T_{3}&=-\sfrac{1}{4}\left(X_{3}^{2}+X_{4}^{2}\right) \label{t3}
\end{align}
which lead us to a quartic algebra depending now explicitly on the Cartan generators.
\begin{align}
[T_{121},T_1]&=2\{ T_{12},T_1 \}-X_2^2 T_{12} \\
[T_{122},T_2]&=2\{ T_{12},T_2 \}-(X_1^2{+}2X_1X_2{+}X_2^2)T_{12} \\
[T_{121},T_2]&=[T_{122},T_1]=-\{ T_{12},T_1 \}-\{ T_{12},T_2 \}+\{ T_{12},T_3 \}+(X_1X_2{+}X_2^2)T_{12} \\
[T_{121},T_3]&=-\{ T_{12},T_1 \}+\{ T_{12},T_2 \}-\{ T_{12},T_3 \}-X_1 X_2 T_{12}  \\
[T_{122},T_3]&=\{ T_{12},T_1 \}-\{ T_{12},T_2 \}-\{ T_{12},T_3 \} +(X_1^2{+}X_1X_2)T_{12} 
\end{align}  
\begin{align} 
[T_{121},T_{12}]&=2\{ T_{122},T_1 \} +\{ T_{121},T_1 \}+\{ T_{121},T_2 \}-\{ T_{121},T_3 \}{-}(X_1X_2{+}X_2^2)T_{121}-X_2^2 T_{122}\\
[T_{122},T_{12}]&=-2\{ T_{121},T_2 \} -\{ T_{122},T_1 \}-\{ T_{122},T_2 \}+\{ T_{122},T_3 \} {+}(X_1^2{+}X_1X_2{+}X_2^2)T_{121}{+}(X_1X_2{+}X_2^2) T_{122}\
\end{align}
\begin{align} 
[T_{121},T_{122}]&= -\{\{ T_{12}, T_1\}, T_1\}{-}\{ \{ T_{12}, T_2\}, T_2\}{-}\{ \{ T_{12}, T_3\}, T_3\}{+}2\{ \{ T_{12}, T_1\}, T_2\}{+}2\{ \{ T_{12}, T_1\}, T_3\}\\
&{+}2\{ \{ T_{12}, T_2\}, T_3\}{+}2(X_1^2{+}X_1X_2) \{ T_1, T_{12} \}{-}2X_1X_2\{ T_2, T_{12} \}{+}2(X_1X_2{+}X_2^2) \{ T_3, T_{12} \}
\end{align}
%
%
%

{}

\end{document}